# Electrodeposition as a new route to synthesize superconducting FeSe

Satoshi Demura,[1,2] Hiroyuki Okazaki,[1] Toshinori Ozaki,[1] Hiroshi Hara,[1,2] Yasuna Kawasaki,[1,2] Keita Deguchi,[1,2] Tohru Watanabe,[1,2] Saleem James Denholme,[1] Yoshikazu Mizuguchi,[1,3] Takahide Yamaguchi,[1] Hiroyuki Takeya,[1] and Yoshihiko Takano[1,2]

[1]National Institute for Materials Science, 1-2-1 Sengen, Tsukuba, Ibaraki 305-0047, Japan

[2]University of Tsukuba, Graduate School of Pure and Applied Sciences, Tsukuba, Ibaraki 305-8577, Japan

[3]Tokyo Metropolitan University, Graduate School of Science and Engineering, Hachioji, Tokyo 192-0397, Japan

**Abstract**

We have successfully synthesized FeSe films by the electrochemical deposition in the electrolyte containing $FeCl_2 \cdot 4H_2O$, $SeO_2$ and $Na_2SO_4$. The composition ratio of Fe and Se was controlled by the synthesis voltage and pH value. The FeSe film with the composition ratio of Fe : Se = 1 : 1 is fabricated at a voltage of -0.9 V and pH 2.1 in our electrochemical deposition. This sample has a highly crystalline tetragonal FeSe structure and exhibits a superconducting transition at 8.1 K, comparable to FeSe synthesized by other methods.

## Introduction

Iron-based superconductors are promising material for superconducting applications under a high magnetic field since they have a high superconducting properties such as high upper critical field $H_{c2}$ [1] and irreversibility field $H_{irr}$ [2]. Among them, iron chalcogenides have some of the simplest structures with a less toxicity than iron pnictides. The iron chalcogenide FeSe with a superconducting transition temperature ($T_c$) of 8 K [3] reaches 37 K under high pressure [4-7]. The large enhancement of $T_c$ is strongly related to the change of the anion height [8]. Therefore, FeSe wires and tapes exhibit high $T_c$ compared to that of the bulk due to the change of the anion height by the shrinkage of *c*-axis [9-11]. From these aspects, the iron chalcogenides can be potential candidates for the application for wires, tapes, and thin films. FeSe thin films have been synthesized by the PLD and MBE techniques. These techniques require expensive equipment and the high vacuum systems. On the other hand, an electrochemical deposition is not only easy, fast and low-cost method to fabricate thin films over large areas compared to PLD and MBE techniques but also easy to apply to tape fabrication and coating. Thus, FeSe electrochemical deposition may be innovative method for the fabrication of superconducting film, wires, tapes, and coatings.

Here, we report the electrochemical synthesis of FeSe films using $FeCl_2{\cdot}4H_2O$, $SeO_2$ and $Na_2SO_4$ as starting materials and the successful for deposition of superconducting FeSe films showing a $T_c$ of 8.1 K.

## Experimental

The electrochemical depositions were performed by a three-electrode method. As an anode, a cathode and a reference electrode, we used a Pt plate, an Fe plate and a Ag/AgCl electrode, respectively. The electrolyte was prepared by dissolving 0.03 mol/l $FeCl_2 \cdot 4H_2O$, 0.015 mol/l $SeO_2$, and 0.1 mol/l $Na_2SO_4$ into distilled water. $H_2SO_4$ was used as a pH adjustor of the electrolyte. We measured cyclic voltammetry (CV) corresponding to the voltage dependence of the current density $J$ within the range from 0 to -3 V vs. Ag/AgCl with the scan rate of 100 mV/s. X-ray diffraction was carried out by the $\theta$-2$\theta$ method with Cu-K$\alpha$ radiation. The composition of the films was determined by the energy dispersive x-ray (EDX) spectrometry. The temperature dependence of magnetization was observed by a superconducting quantum interface device (SQUID) magnetometer with an applied field of 2 Oe.

## Results and discussion

We performed the CV measurement in order to identify a suitable voltage for the deposition of FeSe, as shown in Fig. 1. The cyclic voltammogram shows an anomaly around -0.9 V vs. Ag/AgCl on the forward and reverse scan correspond to the red and blue lines. Since the anomaly probably corresponds to the electrodeposition of FeSe, we performed the electrodeposition in the range between -0.7 and -1.1 V. Figure 2(a) shows the voltage dependence of the composition ratio of Fe and Se. In the film fabricated between -0.9 and -1.0 V, we observed that the Fe and Se composition ratio are comparable, i.e. Fe : Se = 1 : 1. Se composition linearly increases with increasing voltage, while Fe composition decreases. This result indicates that the composition of film can be controlled by the synthesized voltage. The crystal structure of the samples synthesized between -0.7 and -1.1 V are measured by x-ray diffraction, as shown in Fig. 2(b). The film synthesized at -0.9 V shows the highest intensity of tetragonal FeSe peaks than films of other voltages, indicating that the film with the same Fe and Se composition ratio shows good crystallinity of tetragonal FeSe in electrodeposition. Apart from -0.9 V, XRD patterns show broadened peaks of tetragonal FeSe and some evidence of hexagonal Se peaks. From the voltage dependence, we found that the suitable condition to deposit FeSe film is the voltage around -0.9 V in this solution.

Additionally, we investigated the electrodeposited films with changing pH value of the electrolyte. Figure 3(a) shows the pH dependence of the composition ratio of Fe and Se at the voltage of -0.9 V. From the EDX results, we found that the sample with the same Fe and Se composition ratio is obtained around pH 2.1. With increasing pH, Se composition increases while Fe composition decreases. Thus, the composition ratio of the electrodeposited film can be tuned by the pH value. XRD patterns of these films deposited between pH 2.0 and 2.2 are summarized in Fig. 3(b). The XRD pattern of the film synthesized at pH 2.1 clearly shows higher crystallinity of tetragonal FeSe than other pH values. These results indicate that the electrodeposited film with same Fe and Se composition ratio shows well crystallized of tetragonal FeSe. We find that the suitable condition to synthesize FeSe film at the voltage of -0.9 V and the pH 2.1.

Figure 4 shows the temperature dependence of the susceptibility of the obtained sample at suitable condition peeled from Fe substrate was measured after the zero field cooling. The FeSe film with same Fe and Se composition ratio and good crystallinity exhibits the superconducting transition at 8.1 K. The positive magnetization may correspond to Fe particles containing the sample when peeling off the Fe substrate. The $T_c$ is comparable to that of the bulk FeSe. Therefore, the electrochemical deposition can be an innovative method for the fabrication of superconducting films, wires, tapes, and coatings.

Previously, we have published the first report that the superconducting FeSe film was successfully fabricated by an electrochemical syntheses using the electrolyte with $FeSO_4 \cdot 7H_2O$ and $SeO_2$ [12]. However, this FeSe film, synthesized at a voltage of -1.75 V and pH 2.3 shows broader XRD peaks and lower $T_c$ of 3.5 K than the present work. It is expected that the improvement of the film quality is due to the sample synthesizing voltage which minimize the hydrogen generation. Therefore, we succeeded to electrodeposit FeSe film with good crystallinity and higher $T_c$ than previous report.

## Conclusion

We performed the electrochemical synthesis using the electrolyte with $FeCl_2 \cdot 4H_2O$, $SeO_2$ and $Na_2SO_4$ and found that the suitable condition at -0.9 V and pH 2.1. The obtained sample shows the superconducting transition temperature of $T_c = 8.1$ K, which is higher than that of the previous report. Our results demonstrate that the electrochemical deposition can be a powerful technique to easily fabricate superconducting films, wires, tape and coatings.

## Acknowledgements

This work was partly supported by a Grant-in-Aid for Scientific Research from the Ministry of Education, Culture, Sports, Science and Technology (KAKENHI).

**Figure caption**

Fig. 1 Cyclic voltammogram in distilled water dissolving 0.03 mol/l $FeCl_2 \cdot 4H_2O$, 0.015 mol/l $SeO_2$, and 0.1 mol/l $Na_2SO_4$. The red and blue lines correspond to the forward and reverse scan, respectively.

Fig. 2 Voltage dependence of (a) the composition ratio of Fe and Se in the electrodeposited films at pH 2.1 analyzed by EDX and (b) x-ray diffraction patterns of FeSe films fabricated at the voltage between -1.1 and -0.7 V. Peaks marked by ○ and ▲ indicate tetragonal FeSe and hexagonal Se, respectively.

Fig. 3 pH dependence of (a) composition ratio Fe and Se in FeSe films synthesized at the voltage of -0.9 V measured by EDX and (b) x-ray diffraction patterns of FeSe thin films fabricated between pH 2.0 and 2.2 at the voltage of -0.9 V. Peaks denoted by ○ indicate tetragonal FeSe.

Fig. 4 Temperature dependence of the magnetic susceptibility for FeSe film electrodeposited at the voltage -0.9 V and pH 2.1.

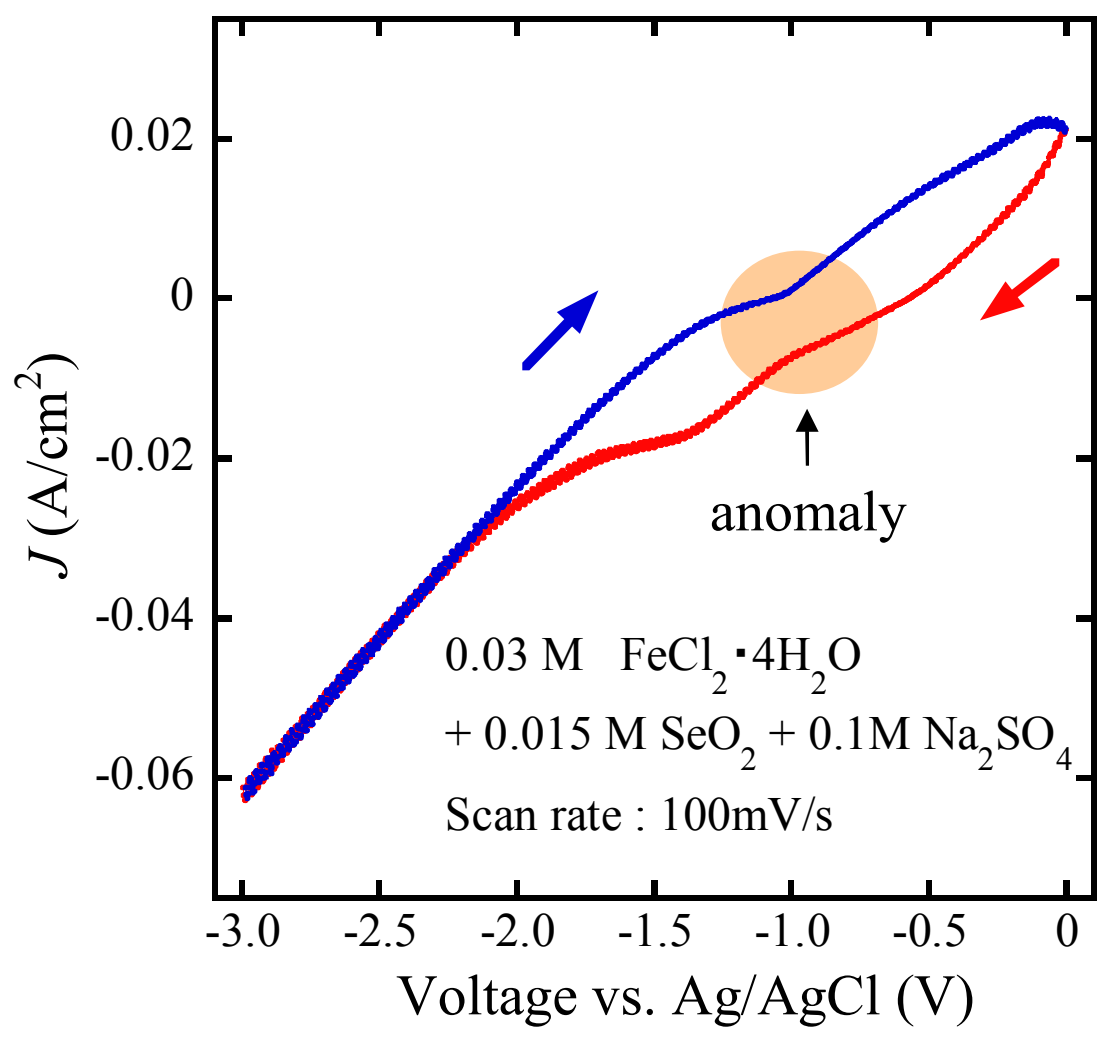

Fig. 1. S. Demura

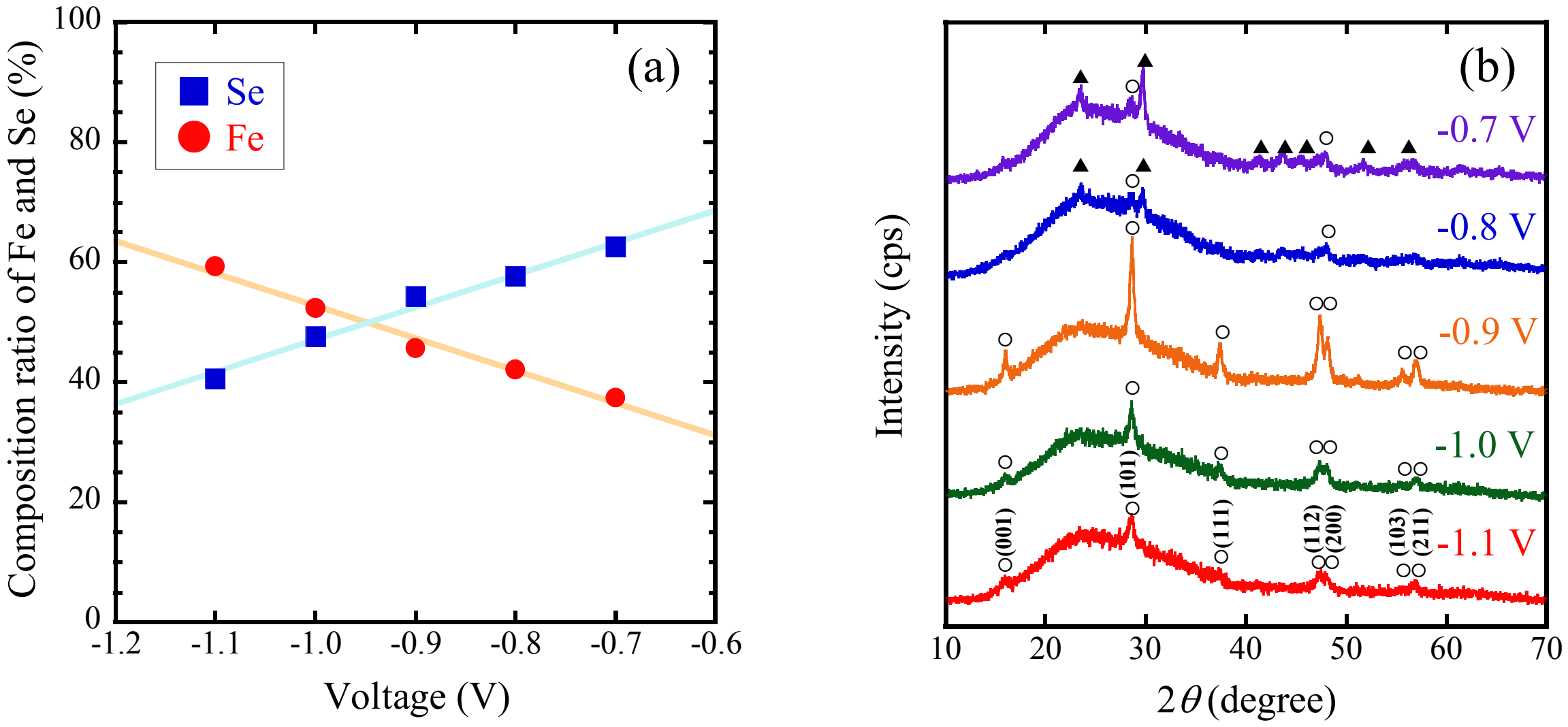

Fig. 2. S. Demura

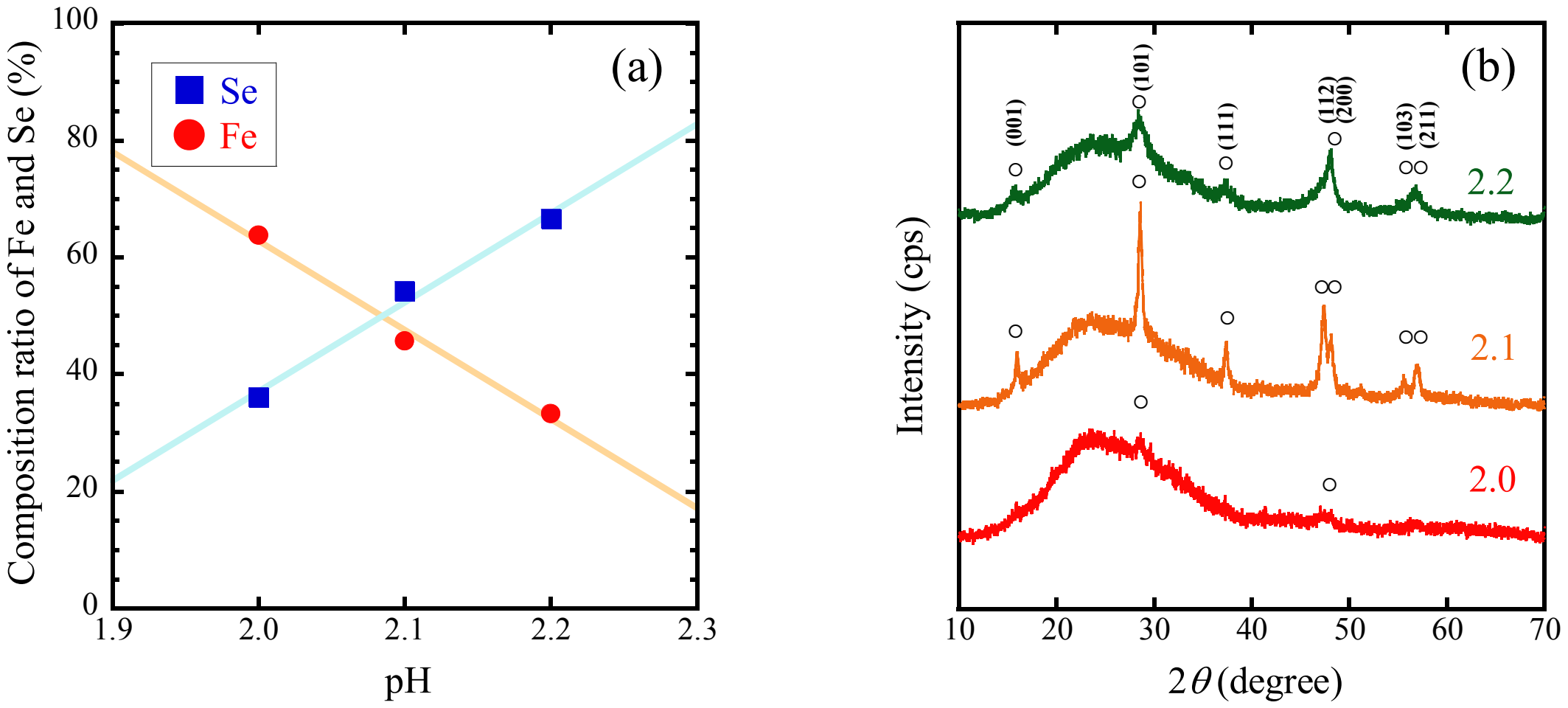

Fig. 3. S. Demura

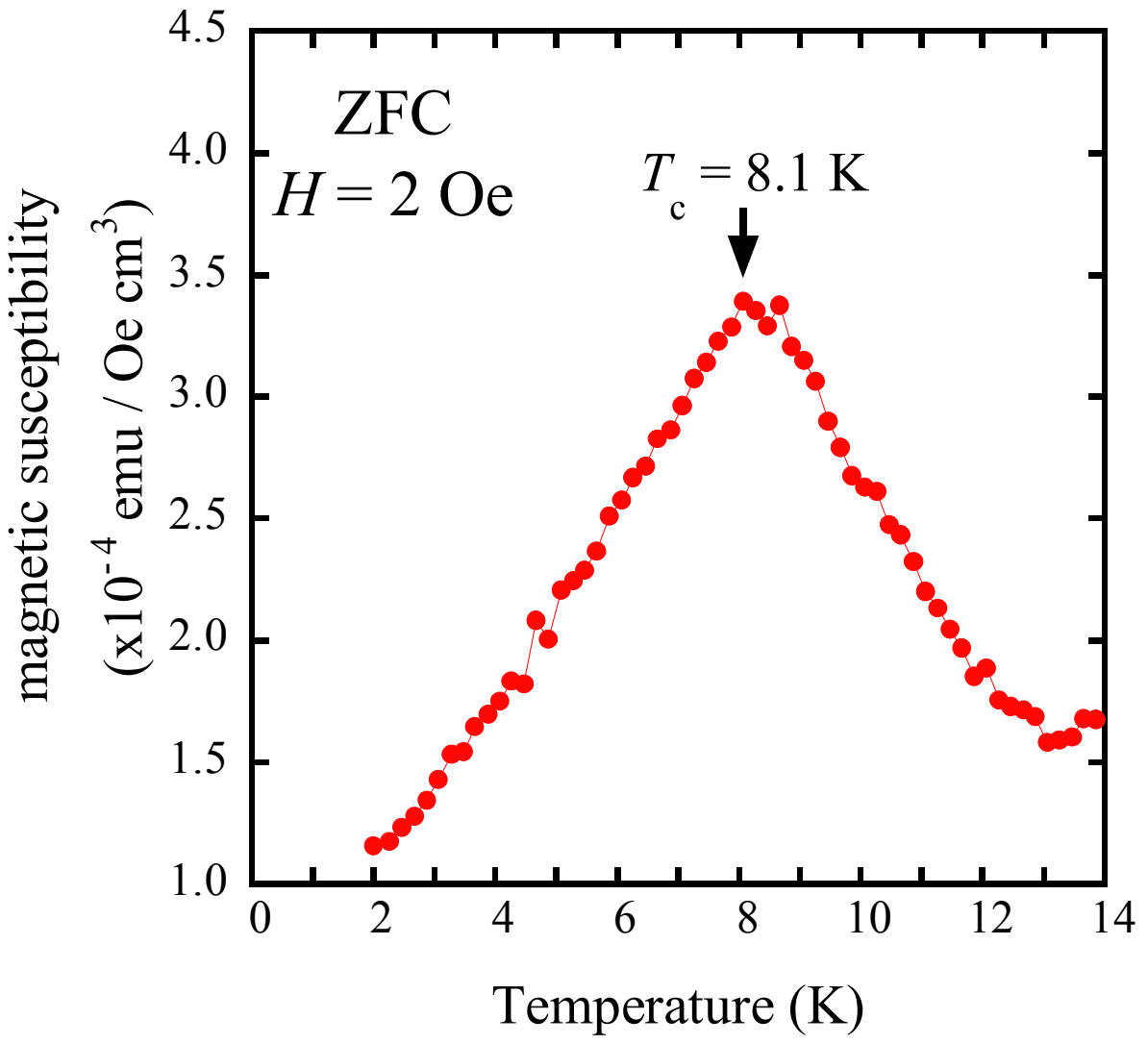

Fig. 4. S. Demura